\documentclass{kluwer}    
\usepackage{psfig}
\newcommand\farcs{\mbox{$.\!\!^{\prime\prime}$}}%

\begin{document}                                                                                   
\begin{article}
\begin{opening}         
\title{Supermassive Black Holes and Galaxy Morphology} 
\author{F.\ D.\ \surname{Macchetto}\thanks{On assignment from the Astrophysics Division, Space Science Department of the European Space Agency}}  
\runningauthor{F.\ D.\ Macchetto}
\runningtitle{Supermassive Black Holes and Galaxy Morphology}
\institute{Space Telescope Science Institute}
\date{\today}

\begin{abstract}
This review addresses one of the important topics of current astrophysical research, namely the role that supermassive black holes play in shaping the morphology of their host galaxies. There is increasing evidence for the presence of massive black holes at the centers of all galaxies and many efforts are directed at understanding the processes that lead to their formation, the duty cycle for the active phase and the question of the fueling mechanism. Related issues are the epoch of formation of the supermassive black holes, their time evolution and growth and the role they play in the early ionization of the Universe. Considerable observational and theoretical work has been carried out in this field over the last few years and I will review some of the recent key areas of progress.
\end{abstract}

\end{opening}

\section{Introduction}  

It is now widely accepted that quasars (QSOs) and Active Galactic Nuclei (AGN) are powered by accretion onto massive black holes. This has led to extensive theoretical and observational studies to elucidate the properties of the black holes, the characteristics of the accretion mechanisms and the mechanisms responsible for the production and transportation of the energy from the central regions to the extended radio lobes. 

However, over the last few years there has been an increasing realization that Massive Dark Objects (MDOs) may actually reside at the centers of \emph{all} galaxies (Ho 1998; Magorrian et~al.\ 1998; Richstone et~al.\ 1998; Van der Marel 1998). The mean mass of these objects, of order $10^{-2.5}$ times the mass of their
host galaxies, is consistent with the mass in black holes needed to produce the observed energy density in quasar light if we make reasonable assumptions about the efficiency of quasar energy production (Chokshi and Turner 1992; Blandford 1999). This raises a number of important new questions and has fundamental implications for the role of the black holes in contributing or being responsible for the ionization (or reionization) of the early universe and for their role in the processes leading to the formation of galaxies. Conversely the apparent correlation between the black hole mass and the mass of the spheroidal component in elliptical and spirals points towards a close interaction between the galaxy size and morphology and its central black hole. Models in which elliptical galaxies form from the mergers of disk galaxies whose bulges contain black holes are consistent with the "core fundamental plane," the relation between the central parameters of early-type galaxies. Furthermore it is clear that the dynamical influence of a supermassive black hole can extend far beyond the nucleus if a substantial number of stars are on orbits that carry them into the center.  Work by Merritt (1998) has shown that nuclear black holes are important for understanding many of the  large-scale properties of galaxies, including the fact that elliptical galaxies come in two, morphologically-distinct families, the absence of bars in most disk galaxies, and the shapes of the spiral galaxy rotation curves. Since the growth of the black hole mass depends on the global morphology of the host galaxies, the link between black hole and galaxy structure implies a feedback mechanism that determines what fraction of a galaxy's mass ends up in the center.

\section{Dynamical Evidence for Massive Black Holes}

\subsection{Megamasers in NGC 4258}

The best observation showing the presence of a Keplerian disk around a black hole was the VLBI observation of megamasers in the nucleus of the Seyfert~2 galaxy NGC~4258 reported by Miyoshi et~al.\ (1995). These observations reveal individual masing knots revolving at distances ranging from $\sim13$~pc to 25~pc around the central object. These data show a near-perfect Keplerian velocity distribution, implying that almost all the mass is located well within the inner radius where the megamasers reside,and they derive a central mass of $\sim3.6\times10^7$~M$_{\odot}$ within the inner $\sim13$~pc. It is not possible to have a cluster of distinct, dark, massive objects (e.g., neutron stars) responsible for such gravitational potential; since the objects would escape on a relatively short time scale, and form a potential well with a different shape, which would force a departure of the megamaser emitting material from pure Keplerian motion (Maoz 1995). Thus the simplest explanation is that the central mass condensation is indeed a black hole.

Given this mass and the fact that NGC~4258 is a relatively low luminosity ($\sim10^{42}$~erg~s$^{-1}$) object, the emission is sub-Eddington, with $L/L_{\rm E}\sim3\times10^{-4}$. Such sub-Eddington sources are likely to have accretion disk structures, where the accreting gas is optically thin and radiates inefficiently, and the accretion energy that is dissipated viscously, is advected with the accretion flow (see, e.g., Ichimaru 1977; Narayan \& Yi 1994; Abramowicz et~al.\ 1995). However, it is important to note that such low Eddington rate cannot be normal among quasars; if most of them radiated at such low $L/L_{\rm E}$, the black hole masses of the most luminous sources would be much larger than expected on other grounds, unless  the low activity is not due to a low black hole mass, but rather to a low accretion rate (Ptak 1997).

\subsection{Kinematic Studies Using Optical Emission Lines}

The other line of evidence for the presence of black holes in galaxies is the velocity field of the matter emitting closely to the nucleus. This approach dates back to the pioneering ground-based observations made in the late 70s, when Sargent and collaborators showed that the stellar velocity dispersion in the radio galaxy M87 increases to 350~km~s$^{-1}$ in the innermost 1.5$''$ from the nucleus.  Considerably better spatial resolution observations using the long-slit spectrograph on the \textit{Faint Object Camera} of \textit{HST} were carried out by Macchetto et~al.\ (1997). They observed the ionized gas disk in the emission line of [OII]$\lambda3727$ at three different positions separated by 0.2 arcsec, with a spatial sampling of 0.03 arcsec (or $\sim2$~pc at the distance of M87). They measured the rotation curve of the inner $\sim1''$ of the ionized gas disk to a distance as close as 0\farcs07 ($\simeq5$~pc) to the dynamical center. They modeled the kinematics of the gas under the assumption of the existence of both a central black hole and an extended central mass distribution, taking into account the effects of the instrumental PSF, the intrinsic luminosity distribution of the line, and the finite size of the slit. They found that the central mass must be concentrated within a sphere whose maximum radius is $\simeq3.5$~pc and showed that both the observed rotation curve and line profiles are best explained by a thin-disk in Keplerian motion. They proved that the observed motions are due to the presence of a super-massive black-hole and derived a value of M$_{\rm BH}= (3.2\pm0.9) \times 10^9$~M$_{\odot}$ for its mass.

\subsection{Virial Masses}

The virial masses and emission-line region sizes of Active Galactic Nuclei (AGNs) can be measured by     ``reverberation-mapping'' techniques. Wandel, Peterson \& Malkan (1999) have compiled a sample of 17 Seyfert~1 and~2 quasars with reliable reverberation and spectroscopic data and used these results to calibrate similar determinations made by photoionization models of the AGN line-emitting regions. Reverberation mapping uses the light travel-time delayed emission-line response to continuum variations to determine the size and kinematics of the emission-line region. The distance of the broad emission-line region (BLR) from the ionizing source is then combined with the velocity dispersion, derived from either the broad H$\beta$ line width or from the variable part of the line profile to estimate the virial mass. When they compare the central masses calculated with the reverberation method to those calculated using a photoionization (H$\beta$ line) model, they find a nearly linear correlation (Table~1). They find that the correlation between the masses is significantly better than the correlation between the corresponding BLR sizes calculated by the two methods, which further supports the conclusion that both methods measure the mass of the central black hole. They also derive the Eddington ratio, which for the objects in the sample fall in the range $L_V/L_{\mathrm{Edd}}\sim0.001$--0.03 and $L_{\mathrm{ion}}/L_{\mathrm{Edd}}\approx0.01$--0.3. 

\begin{kaprotate}
\begin{table}
\caption{Reverberation BLR Sizes and Central Masses Compared with Photoionization Sizes and
Masses. The last two columns give the ionizing luminosity derived from the lag and the
corresponding Eddington ratio. (Wandel et~al.\ 1999)}
\begin{tabular*}{\maxfloatwidth}{lcccclcc}\hline
Name &$\log R_{\mathrm{ph}}$ &log lag &$\log M_{\mathrm{ph}}$ &$\log M_{\mathrm{rev}}$ &\multicolumn{1}{c}{$M_{\mathrm{rev}}(l0^7$ M$_{\odot}$)} &$\log L_{\mathrm{ion}}$ &$\log(L_{\mathrm{ion}}/L_{\mathrm{Edd}})$\\ \hline
3C 120	          &0.92 &1.64 &6.86 &7.49 &3.1$^{+2.0}_{-1.5}$           &45.03 &--0.57\\
3C 390.3         &0.89 &1.38 &8.26 &8.59 &\llap{3}9.1$^{+12}_{-15}$   &44.51	&--2.19\\
Akn 120         &1.07 &1.59 &7.97 &8.29 &\llap{1}9.3$^{+4.1}_{-4.6}$	 &44.92 &--1.48\\
F9                  &1.13 &1.23 &8.03 &7.94 &8.7$^{+2.6}_{-4.5}$            &44.21	 &--1.84\\
1C 4329A &\llap{$<$}0.56 &0.15 &7.34 &\llap{$<$}6.86
&\llap{$<$}0.73~~~ &\llap{$<$}42.04 &\llap{$<$}--2.93\\
Mrk 79	          &0.81 &1.26 &7.48 &8.02 &\llap{1}0.5$^{+4.0}_{-5.7}$ &44.26 &--1.87\\
Mrk 110         &0.78 &1.29 &6.46 &6.91 &0.80$^{+0.29}_{-0.30}$       &44.33	 &--0.69\\
Mrk 335         &0.89 &1.23 &6.68 &6.58 &0.39$^{+0.14}_{-0.11}$       &44.20	 &--0.49\\
Mrk 509         &1.08 &1.90 &7.17 &7.98 &9.5$^{+1.1}_{-1.1}$            &45.54	 &--0.54\\
Mrk 590         &0.85 &1.31 &7.00 &7.15 &1.4$^{+0.3}_{-0.3}$            &44.37	 &--0.89\\
Mrk 817         &0.86 &1.19 &7.53 &7.56 &3.7$^{+1.1}_{-0.9}$            &44.13	 &--1.54\\
NGC 3227      &0.16 &1.04 &6.92 &7.69 &4.9$^{+2.7}_{-5.0}$            &43.82	 &--1.98\\
NGC 3783      &0.52 &0.65 &7.05 &7.04 &1.1$^{+1.1}_{-1.0}$	             &43.05 &--2.10\\
NGC 4051 &\llap{$<$}0.14 &0.81 &5.37 &6.15 &0.14$^{+0.15}_{-0.09}$ &41.57 &--0.84\\
NGC 4151      &0.44 &0.48 &7.35 &7.08 &1.2$^{+0.8}_{-0.7}$            &42.70	 &--2.49\\
NGC 5548      &0.73 &1.26 &7.70 &7.83 &6.8$^{+1.5}_{-1.0}$            &44.27	 &--1.83\\
NGC 7469 &0.90 &0.70 &6.87 &6.88 &0.76$^{+0.75}_{-0.76}$        &43.14	 &--1.86\\
PG 0804+762 &1.39 &2.00 &7.74 &8.34 &\llap{2}1.9$^{+3.8}_{-4.5}$	 &45.75 &--0.70\\
PG 0953+414 &1.54 &2.03 &7.83 &8.19 &\llap{1}5.5$^{+10.8}_{-9.1}$ &45.81 &--0.49\\
\hline
\end{tabular*}
\end{table}
\end{kaprotate}

\subsection{The Black Hole Mass of a Seyfert Galaxy}

In a recent study Winge et~al.\ (1999) have analyzed both ground-based, and \textit{HST/FOC} long-slit spectroscopy at subarcsecond spatial resolution of the narrow-line region (NLR) of NGC~4151. They found that the extended emission gas ($R>4''$) is in a normal rotation in the galactic plane, a behavior that they were able to trace even across the nuclear region, where the gas is strongly disturbed by the interaction with the radio jet and connects smoothly with the large-scale rotation defined by the neutral gas emission. The \textit{HST} data, at 0\farcs03 spatial resolution, allow for the first time truly to isolate the kinematic behavior of the individual clouds in the inner narrow-line region. They find that, underlying the perturbations introduced by the radio ejecta, the general velocity field can still be well represented by planar rotation down to a radius of $\sim0\farcs5$ (30~pc), the distance at which the rotation curve has its turnover. The most striking result that emerges from the analysis is that the galaxy potential derived fitting the rotation curve changes from a ``dark halo'' at the extended narrow-line region distances to being dominated by the central mass concentration in the NLR, with an almost Keplerian falloff in the $1''<R<4''$ interval. The observed velocity of the gas at 0\farcs5 implies a mass of $M\sim10^9$~M$_{\odot}$ within the inner 60~pc. The presence of a turnover in the rotation curve indicates that this central mass concentration is extended. The first measured velocity point (outside the region saturated by the nucleus) would imply an enclosed mass of $\sim5\times10^7$~M$_{\odot}$ within $R\sim0\farcs15$ (10~pc), which represents an upper limit to any nuclear point mass.

\section{Extended Nuclear Disks}

Observations of a number of extended (a few 100~pc) nuclear disks with the \textit{HST} has provided new evidence and constraints on the mass of the MDOs in early type galaxies. 

\subsection{NGC 6251}
Ferrarese and Ford (1999) carried out \textit{HST} imaging and spectroscopy of NGC~6251, a giant E2~galaxy and powerful radio source which is at a distance of $\sim106$~Mpc. The \textit{WFPC2} images at 0\farcs1 (51~pc) resolution, show a well defined dust disk, 730~pc in diameter, whose normal is inclined by $76^{\mathrm{o}}$ to the line of sight. A significant ionized gas component is confined to the central $\sim0\farcs3$ of the disk.  The \textit{FOS} 0\farcs09 square aperture was used to map the velocity of the gas in the central 0\farcs2. They constructed dynamical models assuming two different analytical representations of the stellar mass density and therefore stellar potential, and from the kinematics of the gas they derive a value for the central mass concentration, $4\times10^8$ to $8\times10^8$~M$_{\odot}$. The bolometric non-thermal luminosity of the nucleus is higher than the lower theoretical constraint given by accretion at the Bondi rate and 10\% efficiency which is typical of a conventional thin accretion disk. Therefore this energy output does not require the presence of an advection dominated flow.

Other galaxies studied with \textit{HST} at high spatial resolution include NGC~4261, NGC~4374, NGC~7052 (Ferrarese et~al.\ 1996, Bower et~al.\ 1998, Van der Marel \& Van der Bosch 1998) and show black hole masses in the range 2--$6\times10^8$~M$_{\odot}$.

\subsection{Cen A}
\begin{figure}
\caption{Grayscale representation of the mosaic in the \textit{WFPC2} F814W filter. Surface brightness ranges from 0 (white) to 1.6 in units of $10^{-16}$~erg~s$^{-1}$ cm$^{-2}$~\AA$^{-1}$ arcsec$^{-2}$. Image sizes are $225''\times170''$. North is up and East is left. (Marconi et~al.\ 1999)}
\end{figure}
Centaurus~A (NGC~5128) is the closest giant elliptical galaxy hosting an active galactic nucleus (AGN) and a jet (Figure~1).  Its relative proximity ($D\sim3.5$~Mpc, Hui et~al.\ 1993) offers a unique opportunity to investigate the putative supermassive black hole, the associated accretion disk and jet. However, the study of this nearest giant elliptical at intermediate wavelengths has been severely hindered by the presence of a dust lane which dominated ground-based optical and near-IR observations of the nuclear region. The dust lane, which obscures the inner half kiloparsec of the galaxy, with associated gas, young stars and HII regions, is interpreted as the result of a relatively recent merger event between a giant elliptical galaxy and a small, gas rich, disk galaxy (Baade \& Minkowski 1954, Graham 1979, Malin, Quinn \& Graham 1983).

IR and CO observations of the dust lane have been modeled by a thin warped disk (Quillen et~al.\ 1992; Quillen, Graham \& Frogel 1993) which dominates ground-based near-IR observations along with the extended galaxy emission (Packham et~al.\ 1996). Earlier R-band imaging polarimetry from \textit{HST} with \textit{WFPC} (Schreier et~al.\ 1996) are also consistent with dichroic polarization from such a disk. 

\begin{figure}
\caption{Left panel: contours from the ISOCAM image at 7~$\mu$m. Center Panel: overlay of the ISOCAM contours on the NIC3 Pa$\alpha$ image showing the morphological association between the Pa$\alpha$ emission and the edges of the putative bar. The small right panel is the Pa$\alpha$ disk from Paper~II. Note that its major axis is perpendicular to the edges of the ``bar.'' (Marconi et~al.\ 1999)}
\end{figure}
Recent \textit{HST WFC~2} and \textit{NICMOS} observations of Centaurus~A have shown that the 
20~pc-scale nuclear disk previously detected by NICMOS in Pa$\alpha$ (Schreier et~al.\ 1998) has also been detected in the [FeII]$\lambda1.64~\mu$m line which shows a morphology similar to that observed in Pa$\alpha$ with an [FeII]/Pa$\alpha$ ration typical of low ionization Seyfert galaxies and LINERSs (Figure~2). Marconi et~al.\ (1999) derive a map of dust extinction, E(B--V), in a $20''\times20''$ circumnuclear region and reveal a several arcsecond long dust feature near to but just below the nucleus, oriented in a direction transverse to the large dust lane. This structure may be related to the bar observed with ISO and SCUBA, as reported by Mirabel et~al.\ (1999). They find rows of Pa$\alpha$ emission knots along the top and bottom edges of the bar, with they interpret as star formation regions, possibly caused by shocks driven into the gas. The inferred star formation rates are moderately high ($\sim0.3$~M$_{\odot}$~yr$^{-1}$). If the bar represents a mechanism for transferring gas in to the center of the galaxy, then the large dust lane across the galaxy, the bar, the knots, and the inner Pa$\alpha$ disk all represent aspects of the feeding of the AGN. Gas and dust are supplied by a recent galaxy merger; a several arcminute-scale bar allows the dissipation of angular momentum and infall of gas toward the center of the galaxy; subsequent shocks trigger star formation; and the gas eventually accretes onto the AGN via the 20~pc disk. 

By reconstructing the radial light profile of the galaxy to within 0\farcs1 of the nucleus Marconi et~al.\ (1999) show that Centaurus~A has a core profile. Using the models of van der Marel (1999), they estimate a black hole mass of $\sim10^9$~M$_{\odot}$, consistent with ground based kinematical measurements (Israel 1998).

\section{Statistical Properties of AGN and Radio Galaxies}

An important question about AGN hosts is whether there is anything unusual about their morphology, whether they occur only in a certain type of galaxy, or can be found in all galaxies but their active phase lasts for only a fraction of a Hubble time. Furthermore, the morphology of the host may provide important information about the dynamics that funnel accretion fuel into the nucleus.  Studies of the environments of quasars can also provide insight into the AGN phenomenon in general, such as the relationship between quasars and radio galaxies. If indeed quasars and radio galaxies are objects differentiated only by viewing angle, then quasars might also be expected to exhibit an alignment effect over the same redshift range as the radio galaxies.

The original classification of radio galaxies by Fanaroff \& Riley (1974) is based on a morphological criterion, i.e., edge darkened (FR~I) vs.\ edge brightened (FR~II) radio structure. It was later discovered that this dicothomy corresponds to a continuous transition in total radio luminosity (at 178~MHz) which formally occurs at $L_{178}=2\times10^{26}$~W~Hz$^{-1}$. 

\subsection{Host Morphology of Radio Galaxies and Quasars}

\textit{HST} observations of 273 sources in the 3CR catalog were carried out by Martel et~al.\ (1997, 1999). To study the morphology distribution of the radio galaxies in the sample, they selected those at relatively small redshift. This ensures adequate image quality to permit reliable determination of the morphology, and minimizes the effects due to cosmological evolution of either the population of radio galaxies or the nature of their hosts. The result of this study is that more than 80\% of the radio sources are found in elliptical galaxies, and the remainder have hosts whose morphologies are difficult to determine. 

The 3CR sample is particularly well suited for investigating the relationship between radio galaxies and quasars, and the results have been discussed by Martel et~al.\ (1997, 1999) and Lehnert et~al.\ (1999a) (Figures~3 and~4).

\begin{figure}
\caption{\textit{HST/WFPC2} broad band images of a selection of the 3C sources.}
\end{figure}
\begin{figure}
\caption{\textit{HST/WFPC2} broad band images of a selection of the 3C sources.}
\end{figure}

The study shows that the quasar ``fuzz'' contributes from $<$5\% to as much as 100\% of the total light from the quasar, with a typical value of about 20\%. Most of the sources are resolved and show complex morphology with twisted, asymmetric, and/or distorted isophotes and irregular extensions. In almost every case of the quasars with spatially resolved ``fuzz,'' there are similarities between the radio and optical morphologies. A significant fraction ($\sim25$\%) of the sources show nearby galaxies in projection and $\sim10$\% of the sources show obvious signs of interactions with these nearby companions. These results show that the generally complex morphologies of host galaxies of quasars are influenced by the radio emitting plasma and by the presence of nearby companions.

\begin{figure}
\caption{The host galaxies of three radio loud quasars (Kirhakos et~al.\ 1999).}
\end{figure}
Bahcall et~al.\ (1997) and Kirhakos et~al.\ (1999) have studied in detail nine radio-loud quasars and found that the hosts are either bright ellipticals or occur in interacting systems (Figure~5). There is a strong correlation between the radio emission of the quasar and the luminosity of the host galaxy; the radio-loud quasars reside in galaxies that are on average about 1~mag brighter than hosts of the radio-quiet quasars. 

Further \textit{HST} observations of radio-loud quasars by Lehnert et~al.\ 1999b, analyzed the spatially-resolved structures around five
high-\break redshift radio-loud quasars. 

Comparing the images with high resolution \textit{VLA} radio images they conclude that all of the high redshift quasars are extended in both the rest-frame UV continuum and in Ly$\alpha$. 

The typical integrated magnitude of the host is $V\sim22\pm0.5$, the typical UV luminosity is $\sim 10^{10}$~L$_{\odot}$, and the Ly$\alpha$ images are also spatially-resolved. The typical luminosity of the extended Ly$\alpha$ is about few $\times10^{44}$~ergs~s$^{-1}$; these luminosities require roughly a few percent of the total ionizing radiation of the quasar.

These results show that the generally complex morphologies of host galaxies are influenced by the radio emitting plasma. This manifests itself in the ``alignment'' between the radio, Ly$\alpha$, and UV continuum emission, in detailed morphological correspondence in some of the sources which suggests ``jet-cloud'' interactions, and in the fact that the brightest radio emission and the side of the radio emission with the shortest projected distance from the nucleus occurs on the same side of the quasar nucleus as the brightest, most significant Ly$\alpha$ emission. 

There are few studies of radio-quiet quasars. Some early \textit{HST} observations by Bahcall et~al.\ (1996, 1997)  and more detailed observations by Disney et~al.\ (1995), Boyce et~al.\ (1996, 1998), showed that for a total of about 25~objects the parent galaxies can be either ellipticals or spirals. Overall there are six clear examples of strong ongoing gravitational interaction between two or more galaxies and in 19 other cases close companion objects are detected, suggesting recent gravitational interaction.

\subsection {Seyfert Morphologies}

The study of fueling processes in AGNs is key to our understanding of the structure and evolution of the central black hole and their host galaxies. Although fuel is readily available in the disk, it needs to overcome the centrifugal barrier to reach the innermost regions in disk and elliptical galaxies. Large-scale 
non-axisymmetries, such as galactic bars, are thought to be related to starburst activity within the central kpc, which preferentially occurs in barred hosts (e.g., Heckman 1980; Balzano 1983; Devereux 1987; Kennicutt 1994). In a number of early optical surveys, the fueling of Seyfert activity in disk galaxies was linked to non-axisymmetric distortions of galactic gravitational potentials by large-scale stellar bars and tidal interactions (Adams 1977; Simkin, Su \& Schwarz 1980; Dahari 1985a). This was supported by the  argument that gravitational torques are able to remove the excess angular momentum from gas, which falls inwards, giving rise to different types of activity at the center (Sellwood \& Wilkinson 1993; Phinney 1994). However early studies were not successful in showing significantly higher fractions of bars in host galaxies of AGN. Recently Knapen, Shlosman \& Peletier (1999) have carried out NIR observations at high spatial resolution, on a sample of 34 non-active galaxies from the CF3 catalogue as well as a sample of 48 AGN from the CFA survey. They find that Seyfert hosts are barred more often than normal galaxies, 79\% $\pm$ 7.5\% barred for the Seyferts, vs.\ 59\% $\pm$ 9\% for the control sample, which is 2.5$\sigma$ result.

Their result suggests, but does not prove, that there is an underlying morphological difference between Seyfert and non-Seyfert galaxies, and emphasize the prevalence of barred morphologies in disk galaxies in general, and in active galaxies in particular. 

\subsection{Seyfert Nuclei}

In the standard paradigm where AGNs are powered by non-spherical accretion onto massive black holes, the AGN's luminosity is proportional to the black hole mass accretion rate, which is about 0.01~M$_{\odot}$ year$^{-1}$ for a bright Seyfert nucleus. Strong interactions or mergers with another galaxy are very efficient at funneling large amounts of gas by distorting the galactic potential and disturbing the orbits of gas clouds (Shlosman et~al.\ 1989, 1990, Hernquist and Mihos 1995). This fuel is then brought down to several thousand Schwarzschild radii, or $10^{17}$~cm for a black hole mass of $10^8$~M$_{\odot}$, at which point viscous processes drive the final accretion onto the black hole. 

However, direct observational evidence that galaxy encounters stimulate the luminosity of an AGN has been ambiguous (Adams 1977; Petrosian 1983; Kennicutt \& Keel 1984; Dahari 1985a,  1985b; Bushouse 1986; Fuentes-Williams \& Stocke 1988).

Malkan, Gorjiam and Tam (1998) have recently published the results of an \textit{HST} snapshot imaging survey of  256 cores of active galaxies selected from the ``Catalog of Quasars and Active Nuclei'' by 
V\'eron-Cetty and V\'eron (1986, 1987). Of these, 91 are galaxies with nuclear optical spectra classified as ``Seyfert~1,'' 114 galaxies are classified as ``Seyfert~2,'' and 51 galaxies are classified as ``HIIs.'' This large sample of high-resolution images was used to search for statistical differences in their morphologies. 

The Seyfert galaxies do not, on average, resemble the HII galaxies, which have more irregularity and clumpiness associated with their high rates of current star formation. Conversely, none of the HII galaxies have the filaments or wisps photoionized by the active nucleus which are seen in Seyfert~1 and~2 galaxies.  Of the Seyfert~1 galaxies, 63\% have an unresolved nucleus, 50\% of which are saturated, and 6\% have such dominant nuclei that they would appear as ``naked quasars'' at higher redshifts. The presence of an unresolved and/or saturated nucleus is anti-correlated with an intermediate spectroscopic classification (such as Seyfert~1.8 or 1.9) and implies that those Seyfert~1s with weak nuclei in the \textit{HST} images are extinguished and reddened by dust.

The vast majority of the Seyfert~2 galaxies show no central point source. If all Seyfert~2s were to have unresolved continuum sources like those in Seyfert~1s, they would be at least an order of magnitude fainter. In those galaxies without any detectable central point source (37\% of the Seyfert~1s; 98\% of the Seyfert~2s, and 100\% of the HIIs), the central surface brightnesses are statistically similar to those observed in the bulges of normal galaxies.

Seyfert~1s and 2s both show circumnuclear rings in about 10\% of the galaxies. Malkan et~al.\ (1998) identified strong inner bars as often in Seyfert~1 galaxies (27\%) as in Seyfert~2 galaxies (22\%).

The Seyfert~2 galaxies are more likely than Seyfert~1s to show irregular or disturbed dust absorption in their centers as well as galactic dust lanes which pass very near their nuclei, and on average, tend to have latter morphological types than the Seyfert~1s. Thus it appears that the host galaxies of Seyfert~1 and~2 nuclei are \emph{not} intrinsically identical. A galaxy with more nuclear dust and in particular more irregularly distributed dust is more likely to harbor a Seyfert~2 nucleus. This indicates that the higher dust-covering fractions in Seyfert~2s are the reason for their spectroscopic classification: their compact Seyfert~1 nucleus may have been obscured by galactic dust. This statistical result does not agree with the unified scheme for Seyfert galaxies, thus Malkan et~al.\ (1998) propose that the obscuration which converts an intrinsic Seyfert~1 nucleus into an apparent Seyfert~2 occurs in the host galaxy hundred of parsecs from the nucleus. If so, this obscuration may have no relation to a hypothetical dust torus surrounding the central black hole.

\subsection{Morphologies of FR I Radio Galaxies}

Significant progress in the understanding of the inner structure of FR~I have been obtained thanks to \textit{HST} observations. A newly discovered feature in FR~I are faint, nuclear optical components, which might represent the elusive emission associated with the AGN. Their study can be a powerful tool to directly compare the nuclear properties of FR~I with those of other AGNs, such as BL~Lac objects and powerful radio galaxies.

Chiaberge, Capetti \& Celotti (1999) have studied a complete sample of 33 FR~I sources from the 3CR observations carried out as part of the \textit{HST} snapshot survey and discussed by Martel et~al.\ (1997, 1999) (objects with $z<0.1$) and by de~Koff et~al.\ (1996) (objects with $0.1<z<0.5$).  Chiaberge et~al.\ (1999) have shown that an unresolved nuclear source (Central Compact Core, CCC) is present in the great majority of these objects. The CCC emission, found to be strongly connected with the radio core emission, is anisotropic and can be identified with optical synchrotron radiation produced in the inner regions by a relativistic jet. These results are qualitatively consistent with the unifying model in which FR~I radio galaxies are misoriented BL~Lacs objects.  However, the analysis of objects with a total radio power of $<2\times10^{26}$~W~Hz$^{-1}$, shows that a CCC is found in all galaxies except three, for which absorption from extended dust structures clearly plays a role. This result casts serious doubts on the presence of obscuring thick tori in FR~I as a whole.

The CCC luminosity represents a firm upper limit to any thermal component, and implies an optical luminosity of only $\lsim\ 10^{-5}$--$10^{-7}$ times Eddington (for a $10^9$~M$_{\odot}$ black hole). This limit on the radiative output of accreting matter is independent from but consistent with those inferred from X-ray observations for large elliptical galaxies, thus suggesting that accretion might take place in a low efficiency radiative regime (Fabian \& Rees 1995).

The picture which emerges is that the innermost structure of FR~I radio galaxies differs in many crucial aspects from that of the other classes of AGN; they lack the substantial BLR, tori and thermal disk emission, which are usually associated with active nuclei. Similar studies of higher luminosity radio galaxies will be clearly crucial to determine if either a continuity between low and high luminosity sources exists or, alternatively, they represent substantially different manifestations of the accretion process onto a supermassive black hole.

\section{Demographics of Massive Black Holes}

\subsection{Ellipticals and S0 Galaxies}

The evidence that massive dark objects (MDOs) are present in the centers of nearby galaxies has been reviewed by Kormendy \& Richstone (1995), Bender, Kormendy, \& Dehnen (1996), Kormendy et~al.\ (1997), and van der Marel (1998). The MDOs are probably black holes, since star clusters of the required mass and size are difficult to construct and maintain, and since black hole quasar remnants are expected to be common in galaxy centers. Kormendy \& Richstone suggest that at least 20\% of nearby kinematically hot galaxies (ellipticals and spiral bulges) have MDOs and show a correlation $M_{\bullet} \simeq 0.003 M_{\mathrm{bulge}}$, where $M_{\mathrm{bulge}}$ is the mass of the hot stellar component of the galaxy. For a ``bulge'' with constant mass-to-light ratio $\Upsilon$ and luminosity $L$, $M_{\mathrm{bulge}} \equiv\Upsilon L$.

\begin{figure}
\caption{(a)~Observed black hole masses and bulge luminosities for the samples of nearby galaxies compiled by Ho 1998, \textit{circles}) and Magorrian et~al.\ (1998, \textit{squares}). The solid line shows the linear least square fit to the data, $\log(M_{\mathrm{BH}}$~M$_{\odot})=1.28\log
(L_{\mathrm{B}}$/L$_{\odot})-4.46$, while the dashed lines show the $\pm1\sigma$ deviation ($\sigma =0.74$). The dashed lines are the same in all panels for comparison. (b)~Monte Carlo simulations of black hole masses and bulge luminosities with $M_{\mathrm{acc}}=6\times10^{-3}\Delta M_{\ast}$. (c)~Same as (b) with $M_{\mathrm{acc}}=1.4\times10^{-3}(1+z)^2\Delta M_{*}$. (d)~Same as (b) with $M_{\mathrm{acc}}=10^{-6}(1+z)^2\ M_{\mathrm{halo}}\ \exp[-v_c/300$~km~s$^{-1})^4]$. The linear least square fit to the results of model (iii) has a slope of 0.6, shown by the two dotted lines (Cattaneo et~al.\ 1999).}
\end{figure}
To further probe this correlation Magorrian et~al.\ (1998) constructed dynamical models for a sample of 36 nearby galaxies with \textit{HST} photometry and ground-based kinematics. The models assume that each galaxy is axisymmetric, with a two-integral distribution function, arbitrary inclination angle, a 
position-independent stellar mass-to-light ratio $\Upsilon$, and a central massive dark object (MDO) of arbitrary mass $M_{\bullet}$.  They provide acceptable fits to 32 of the galaxies, and the mass-to-light ratios inferred show a correlation $\Upsilon\propto L^{0.2}$ (Figure~6). 

The result is that virtually every hot galaxy hosts a MDO with a mass ranging from $\sim10^8$~M$_{\odot}$ to $2\times10^{10}$~M$_{\odot}$ and roughly proportional to the mass of the spheroidal stellar component  $M_{\bullet}\sim 0.006 M_{\mathrm{bulge}}$. MDO masses are just large enough to match those related to the QSO phenomenon. In fact, the highest bolometric luminosities of Quasars ($L_{\mathrm{bol}}\ \lsim\ 4\times10^{48}$~erg/s) imply, under the assumption that they radiate at the Eddington limit, that the underlying black hole masses are comparable with those of the biggest MDOs detected in ellipticals, while the lowest QSO bolometric luminosities $L_{\mathrm{bol}}=10^{46}$~erg/s still imply quite conspicuous black hole masses: $M_{\mathrm{BH}}>2\times10^8$~M$_{\odot}$.

\subsection{Early and Late Type Spirals}

Salucci et~al.\ (1999) have studied the rotation curves of about one thousand spiral galaxies to investigate whether they could host \emph{relic} black holes. The sample comprises late type spirals with at least one velocity measurement inside 250~pc for 435 objects and inside 350~pc for the remaining $\sim500$ objects. This would allow detections of MDOs of mass $M_{\mathrm{MDO}}\ \gsim\ (1-2)\times 10^8$~M$_{\odot}$, typical of a black hole powering a QSO and much larger than the ordinary stellar component inside this radius.

\begin{figure}
\caption{The upper limits on the MDO mass as a function of luminosity. The solid line represents the corresponding luminosity-averaged values. Also shown the \textit{minimum} mass for QSO remnants (dashed line) (Salucci et~al.\ 1999).}
\end{figure}
The upper limits obtained are shown in Fig.~7: the central objects in spirals are remarkably less massive than those detected in ellipticals: strict upper limits to their mass range are between $10^6$~M$_{\odot}$ at $L_{\mathrm{B}}\simeq1/20L_{\ast}$ to $\simeq10^7$~M$_{\odot}$ at $\sim3L_{\ast}$.

Salucci et~al.\ (1999) also analyzed Sa galaxies which are considerably fewer than late type spirals, but in view of their very massive bulge, $M_b>10^{10}$~M$_{\odot}$, they may well be the location of black holes with $M_{\mathrm{BH}}\ \gsim\ 10^8$~M$_{\odot}$.

They derived MDO upper limits which range from $5\times10^6$~M$_{\odot}<M^{\mathrm {u}}_{\mathrm {MDO}}<10^{10}$~M$_{\odot}$, and are, at a given luminosity, one order of magnitude larger than the upper limits of the late type spirals MDOs. This implies that inside the innermost kpc, early type spirals have a large enough mass to envelop a large MDO and thus comfortably hide the remnant of a bright quasar.

\section{Formation and Evolution}

As we have seen in the previous section there is a strong empirical relationship between black holes and their host galaxies, it is therefore important to compare their properties and distribution during the quasar era at $z\sim3$. Were todays MBHs already fully formed by that time, or was the average MBH smaller in the past and later grew through accretion or mergers to form the present population?

The epoch of maximal activity in the Universe peaked just before the epoch of maximal star formation, and MBHs must have been formed and active before this time to provide the energy to power the quasars. While the rise of luminous quasars follows closely the rise in starbirth, the bright quasars reach their peak at $z\ \gsim\ 2(t\ \lsim\ 1.6\times10^9$~yr), and then their number decline about $10^9$~yr before the peak in star formation, which occurs at $z\sim1.2(t=2.6\times10^9$~yr).

This scenario favors models in which the black hole forms before the formation of the densest parts of galaxies. For this reason we can associate the birth of quasars with the spheroid formation, a process which is closely coupled to dense regions that collapse early.

The decline in the number of quasars at $z<2$ may be caused by several mechanisms including the exhaustion of the available fuel. Galaxy mergers, which are an effective gas transport process, became less frequent as time evolves and involve a lower mean density.

Limits to the total mass of the black hole can occur through different mechanisms. Sellwood \& Moore (1999) have suggested that strong bars form in the centers of recently formed galaxies and channel mass inwards to the central black hole which grows until its mass is $\sim0.02$ of the mass of the disk. The bar then weakens and infalling mass forms a much more massive bulge which creates an inner Lindblad resonance which suppresses re-formation of a bar. Another mechanism proposed by Merritt (1998) is that the black hole makes the central stellar orbits become chaotic, with the consequence that non-axisymmetric disturbances are smoothed out and the rate of infall of accreting gas falls. 

Cattaneo (1999) has investigated a very simple model in which both spheroids and supermassive black holes form through mergers of galaxies of comparable masses. He assumed that cooling only forms disk galaxies and that, whenever two galaxies of comparable masses merge, the merging remnant is an elliptical galaxy, a burst of star formation takes place and a fraction of the gas in the merging remnant is accreted by a central supermassive black hole formed by the coalescence of the central black holes in the merging galaxies.

He found that this simple model is consistent with the shape of the quasar luminosity function, but he also found that its redshift evolution cannot be explained purely in terms of a decrease in the merging rate and of a decline in the amount of fuel available. To explain the evolution of the space density of bright quasars in the interval $0<z<2$, additional assumptions are needed, such as a redshift dependence of the fraction of available gas accreted or of the accretion time-scale. 

In another scenario, proposed by Silk \& Rees (1998) and by Haehnelt et~al.\ (1998), a $\sim10^6$~M$_{\odot}$ black hole forms by coherent collapse in the nucleus before most of the bulge gas turns into stars. If the black hole accretes and radiates at the Eddington limit, it can drive a wind with kinetic luminosity $\sim0.1$ of the radiative luminosity. This deposits energy into the bulge gas, and will unbind it on a dynamical timescale with the result that the black hole mass will be limited to a value where it is able to shut off its own fuel supply (Blandford, 1999).

A bright AGN may also limit infall of gas to form a disk, through Compton heating, radiation pressure on dust or direct interaction with a powerful wind. When the black hole mass and luminosity are large, the weakly bound, infalling gas will be blown away and an elliptical galaxy will be left behind. Only when the black hole mass is small, will a prominent disk develop. In this case, the bulge to disk ratio should correlate with the black hole mass fraction.

In summary, it seems very likely that black holes form first at quite large redshifts, $z\gg2$ and can grow to their present sizes with standard radiative efficiency, by the time of the main quasar epoch at $t\sim3$~Gyr. There are several plausible mechanisms to limit the growth of the black hole by switching off the fuel supply, all of which need to be need to be studied further.

\section{Where are the Central Engines of QSOs Today?}

The results described in the previous sections are consistent with the general black hole interpretation of nuclear activity.

However, several important pieces of this picture are still missing. One is the marked dichotomy in objects presumed to harbor a nuclear black hole. At high redshifts we find strong evidence of activity associated with supermassive black holes, but no direct proof of them, mostly because the search for a black hole requires resolving the radius of influence of the black hole, and this is possible only out to a few tens of Mpc at best. In local objects we showed in \S\S 2.1 and 2.2 good dynamical evidence for the presence of nuclear black holes, but it is unrelated to the AGN activity seen at high~$z$.

Other missing details include the formation and evolution of supermassive black holes and their relationship with the host galaxy population in general. Specific predictions of physical models of quasars and quasar evolution (e.g., Soltan 1982; Chokshy \& Turner 1992; Cavaliere \& Padovani 1989) suggest that the AGN phenomenon is short-lived and probably recurrent, with the implication that a significant fraction of all massive galaxies should contain a black hole of typically $\sim10^8$--$10^9$~M$_{\odot}$, which is currently inactive.

Further arguments in favor of a close relationship between quasar activity and the formation of a massive spheroidal galaxy have been given by Hamann \& Ferland (1993) and Franceschini \& Gratton (1997), based on the chemical properties of the enriched gas in the vicinity of the QSO. These arguments naturally imply that most spheroidal galaxies today should harbor a supermassive black hole in their nuclei. This raises a potential problem for the black hole interpretation, since in elliptical galaxies we would expect considerably more X-ray emission by 
Bondi-accreted halo plasma than is observed (Fabian \& Canizares 1988).

This can be solved invoking advection-dominated accretion flow (ADAF: Rees 1982; Begelman 1986; Narayan \& Yi 1995; Narayan 1995, 1996) occurring at low accretion regimes, or even better with the adiabatic 
inflow--outflow solutions (ADIOS) developed by Blandford \& Begelman (1999). In such a case the radiative efficiency of the accreting material is low, and the energy released by viscous friction is advected into the black hole rather than radiated. Massive black holes would then be very faint at the frequencies at which typical AGNs with higher accretion rates are very prominent.

\begin{figure}
\caption{The mass function of black hole remnants (from Salucci et~al.\ 1998) compared with the Sa and Sb-Im ULMFs. The dots represent the black hole remnants in QSOs, the solid line the MDOs in ellipticals. Solid and hollow triangles indicate ULMFs for Sa and Sb-Im, respectively.}
\end{figure}

Where are the engines of the QSO phase, i.e., where are the masses up to $\sim10^{10}$~M$_{\odot}$ black hole remnants, presently located? While the QSO/AGN phenomenon implies the existence of a large number of black holes with a wide range of masses, $\sim10^7$--$2\times10^{10}$~M$_{\odot}$, the inner kinematics of late type spirals shows that these systems can be major hosts of black holes with mass only in the range 
$10^6$--$10^7$~M$_{\odot}$. Outside these mass values, other Hubble types must provide the locations for the great majority of black hole remnants. The role of the Sa is likely to be statistically significant in the mass range $10^7$--$10^8$~M$_{\odot}$. At higher mass values the mass function of detected MDOs in E and S0 galaxies matches that of the QSO black holes, leaving very little room to  the contribution of an additional component. 

The role of spirals as black hole hosts can be probed by deriving the MDO upper limits mass function obtained by convolving the spirals luminosity functions with the MDO mass limits as a function of the luminosity.
Salucci et~al.\ (1999) have derived an upper limit for the cosmological density of MDOs residing in spirals (Figure~8); assuming that galaxies distribute according to Schechter luminosity functions $\phi(L_{\mathrm{B}},T)$, they get:
\[\rho_{\mathrm{BH}}(Sb-Im)<4.5\times10^4~\mathrm{M}_{\odot}~\mathrm{Mpc}^{-3}\]
\[\rho_{\mathrm{BH}}(Sa)<1.6\times10^6~\mathrm{M}_{\odot}~\mathrm{Mpc}^{-3}\]
These limits should be compared with the cosmological black hole mass density, estimated to be $\rho_{\mathrm{BH}} \simeq 6\times10^5~\mathrm{M}_{\odot}~\mathrm{Mpc}^{-3}$. We can conclude that the contribution to the luminous phase of quasar due to black holes hosted in late type spirals is totally negligible, whereas in addition to ellipticals Sa nuclei could be the location of dormant black holes, once at the heart of the QSO luminous phase.

\acknowledgements
I would like to thank David Axon for a careful reading of the manuscript.

\end{article}
\end{document}